\documentclass[aps,pra,showpacs,twocolumn]{revtex4}
\usepackage{amssymb}
\usepackage{epsfig}
\usepackage{graphicx}
\usepackage{amsmath}
\usepackage{array,color}
\usepackage{mathrsfs}
\usepackage{subfigure}
\begin{document}
\title{Generation of ring dark solitons by phase engineering and their oscillations
in spin-$1$ Bose-Einstein condensates}
\author{Shu-Wei Song$^{1}$, Deng-Shan Wang$^{2}$, Hanquan Wang$^{3}$ and W. M. Liu$^{1}$}
\address{$^1$Beijing National Laboratory for Condensed Matter Physics,
Institute of Physics, Chinese Academy of Sciences, Beijing 100190,
China}
\address{$^2$School of Science, Beijing Information Science and Technology University, Beijing 100192, China}
\address{$^3$School of Statistics and Mathematics, Yunnan University of Finance and Economics,
Kunming, Yunnan Province, 650221, China}
\date{\today}

\begin{abstract}
The ring dark solitons in spin-$1$ $^{23}$Na and $^{87}$Rb Bose-Einstein condensates are studied numerically in the framework of the time-dependent Gross-Pitaevskii equations. By simulating the phase engineering technique in real experiments, we explore the roles of the parameters characterizing the far-off resonant laser pulse which can be used to generate the ring dark solitons. The variations of these parameters have dramatic effect on the lifetime and the decay profiles of the ring dark solitons. If only one ring dark soliton is generated in one component of the condensate, ring dark solitons in other components are inclined to be induced, resulting in a coexistence state composed of interdependent ring dark solitons coming from different components of the condensate. Ring dark solitons in this coexistence state exhibit dynamical oscillations for hundreds of milliseconds. By studying the lifetime and decaying profiles of ring dark solitons, we explore the similarities and differences of $^{23}$Na and $^{87}$Rb condensates. Besides, taking into account the fact that the center of the ring may not be coincide with that of the trap, we study the dynamics and decaying profiles of the off-centered ring dark solitons in the presence of symmetry breaking effect.
\end{abstract}

\maketitle
\section{INTRODUCTION}
Since the experimental realization of the Bose-Einstein condensate (BEC), intensive interest has been growing and the condensate has been a popularly investigated platform for various effects of quantum many-body interaction. This opens a new field of exploring various types of excitations such as bright solitons, dark solitons and vortices \cite{Matthews,Strecker,Burger}. The concept of ring dark soliton was introduced theoretically and realized experimentally in nonlinear optics \cite{Kivshar1994,Nistazakis,Neshev}. Theocharis et al. investigated the ring dark solitons in scalar BEC with disk-shaped trap \cite{Theocharis,Theocharis1,Yang,Frantzeskakis,Hu}. Recently, dark-bright stripe solitons and ring dark solitons in two-component BEC are studied and the stable ring dark-bright solitons are found \cite{Stockhofe,Valeriy}. The deep ring dark soliton in scalar BEC has been found always unstable towards the formation of vortices. However, it is shown in \cite{Stockhofe} that the presence of the ``bright" component has a
stabilizing effect on the symbiotic state.

As far as we know, ring dark solitons in spin-$1$ BEC have not been investigated theoretically or experimentally yet. Spinor BEC liberates the spin degrees of freedom of the atoms confined in optical traps. Several phases are possible below the transition temperature $T_c$, and this kind of diversity leads to a fascinating area of research for quantum gases. The magnetic moment accompanying the spin gives rise to magnetism, resulting in kinds of excitations like spin domains, spin textures and fractional vortices \cite{Stenger,Leanhardt}. In addition, matter-wave dark solitons can be created experimentally by means of various methods, including the phase engineering and density engineering techniques \cite{Dobrek,Denschlag,Dutton,Ginsberg,Shomroni}. Especially, Wright et al. \cite{Wright} have achieved spatially varying control of the amplitude and phase of the spinor order parameter. To shade light on the manipulation of ring dark solitons in spinor BEC, we investigate the generations and oscillations of the ring dark solitons in spin-$1$ BEC by simulating the phase engineering technique.

In the present paper, we explore the dynamical generation and evolution of the ring dark solitons in $F=1$ $^{23}$Na and $^{87}$Rb condensates. By using models to simulate the phase engineering technique in the experiments, we study the effect of two key parameters of the far-off resonant laser pulse on the dynamics of the ring dark solitons, the width of the laser-induced Stark-shift potential and the duration of the laser beam. We study the generation and dynamics of the
coexistence state of ring dark solitons that is induced by producing only one ring dark soliton in one component of the condensate.  Numerical results are compared between $^{23}$Na and $^{87}$Rb condensates.
%It should be noticed that as the RDSs state is not stationary solutions of the Gross-Pitaevskii equations, we use ``relatively stable" to mean the ring dark profiles of the RDSs taking no account of the change of their width and depth for a lifetime of milliseconds.

The rest of the paper is organized as follows. In Sec. \ref{sec1}, we introduce the mean field theoretical
model of spin-$1$ BEC, and a phase distribution model to imitate the phase configuration of the
order parameter after the illumination of the laser beam.
In Sec. \ref{sec3}, we obtain the ground states of the spin-$1$ $^{23}$Na and $^{87}$Rb condensates
numerically, then study the generation process of the ring dark solitons and investigate the roles of
parameters characterizing the far-off resonant laser pulse. Sec. \ref{sec4} focuses on the emergence,
oscillation and decay of the coexistence state of ring dark solitons that is
induced by the ring dark soliton in one component of the condensate. Finally, the conclusions are summarized in
Sec. \ref{sec5}.

\par
\begin{figure}[tbp]
\includegraphics[width=6cm,height=4cm,]{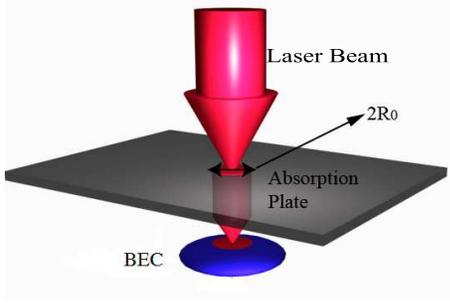}
\caption{(Color online) A schematic of imprinting  phase steps  onto the condensate. The far-off resonant
laser pulse (light red)
goes through an absorption plate (gray), which is used to modulate the laser. The modulated laser pulse
illuminates
the cold atomic condensate (blue) and produces an expected phase configuration. Here, $R_0$ is the radius
of the ring dark soliton
that is to be created. }
\label{fig1}
\end{figure}

\section{The Theoretical Model}\label{sec1}
Most BECs of atomic gases have internal degrees of freedom originating from the spin. For the $^{87}$Rb
and $^{23}$Na atoms, there are spin combinations with $S=1/2$ (electronic spin) and $I=3/2$ (nuclear spin), so the
total spin $F$ can be chosen to be $F=1$ or $F=2$. Both spin-$1$ and spin-$2$ atomic condensates are investigated
experimentally and theoretically \cite{Tetsuo,Stenger,Nistazakis2008,Erich,Bao2010,Ieda,Piotr,Beata,Barnett,
Saito,Nicholas}. For $F=1$, there are three sub-level states with magnetic quantum number $m=0, \pm1$.
Within the mean field description, the spin-$1$ cold atom condensate can be described by a vectorial
order parameter ${\Psi}=(\psi_{1}, \psi_{0}, \psi_{-1})^T$ (the superscript $T$ stands for the transpose), where
the component $\psi_m$ $(m=0,\pm1)$ denotes the macroscopic wave function of the atoms condensed in the hyperfine
state $|F=1, m\rangle$. The dynamics of the vectorial order parameter follows the multicomponent Gross-Pitaevskii equations (GPEs)\cite{Ho,Tetsuo}
\begin{small}
\begin{eqnarray}
i\hbar\frac{\partial\psi_{1}}{\partial t} & = &(-\frac{\hbar ^{2}}{2M}\nabla^{2}+V)\psi _{1}+(c_0+c_2)
(|\psi_{1}|^2+|\psi_{0}|^2)\psi_{1}\notag\\
&&+(c_0-c_2)|\psi_{-1}|^2\psi_{1}+c_2\psi_{0}^2\psi_{-1}^{\ast},\notag\\
i\hbar\frac{\partial\psi _{0}}{\partial t} & = &\big(-\frac{\hbar ^{2}}{2M}\nabla^{2}+V\big)\psi _{0}+(c_0+c_2)
\big(|\psi_{1}|^2+|\psi_{-1}|^2\big)\psi_{0}\notag\\
&&+c_0|\psi_{0}|^2\psi_{0}+2c_2\psi_{1}\psi_{0}^{\ast}\psi_{-1},\notag\\
i\hbar\frac{\partial \psi _{-1}}{\partial t} & = &\big(\!-\!\frac{\hbar ^{2}}{2M}\nabla ^{2}\!+\!V\big)\psi _{-1}\!+\!(c_0+c_2)\!\big(|\psi_{-1}|^2\!+\!|\psi_{0}|^2\big)\psi_{-1}\notag\\
&&+(c_0-c_2)|\psi_{1}|^2\psi_{-1}+c_2\psi_{0}^2\psi_{1}^{\ast},\label{1}\label{GP1}
\end{eqnarray}
\end{small}

\begin{figure}[tbp]
\includegraphics[width=8cm]{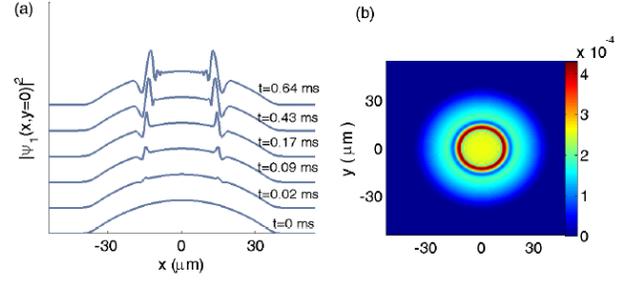}
\caption{(Color online) (a) The 1D density distribution $|\psi_1(x,y=0)|^2$ showing the emergence
of a ring dark soliton in the $\psi_1$ component of the $^{23}$Na condensate with a total number of atoms
$N_{Na}=10^5$, $\Delta x=0.54$ $\mu m$, $\Delta\phi=\pi$ and $R_0=15.77$ $\mu m$. (b) The ring dark soliton configuration in the $\psi_1$ component of the $^{23}$Na condensate at $t=0.64$ $ms$. For the $^{87}$Rb condensate, the process of generating and configuration of the ring dark soliton show no significant difference comparing to the $^{23}$Na condensate.}
\label{fig2}
\end{figure}

\noindent
where $\nabla^2=\partial^2/\partial x^2+\partial^2/\partial y^2+\partial^2/\partial z^2$,
$c_{0} = 4\pi \hbar ^{2}(a_{0}+2a_{2})/3M$ and $c_{2}=4\pi\hbar ^{2}(a_{2}-a_{0})/3M$ ($M$ is the
atomic mass). $a_0$ and $a_2$ are the $s$-wave scattering lengths corresponding to total spin of the two colliding
bosons $0$ and $2$, respectively. The ground state of the condensate is ferromagnetic for $c_2 < 0$ and antiferromagnetic for $c_2 > 0$. $V=M(\omega_x^2x^2+\omega_y^2y^2+\omega_z^2z^2)/2$ ($\omega_{x,y,z}$ is the confinement frequency in the corresponding direction) is the external potential. The density of particles is defined by $n=\sum_m |\psi_m|^2(m=0,\pm1)$ and the total particles number $N=\int n d\vec{r}$.

To investigate the ring dark solitons in a pancake condensate, we choose $\omega_z \gg \omega_x=\omega_y$, and
the 3D GPEs (\ref{GP1}) can be reduced to 2D GPEs. After the dimensionless process (we denote the order parameter with the same $\psi_{m}$), we have
\begin{small}
\begin{eqnarray}
i\frac{\partial \psi _{1}}{\partial t} &=&\big[-\frac{1}{2}(\frac{\partial^2}{\partial x^2}+
\frac{\partial^2}{\partial y^2})+\frac{1}{2%
}\Omega ^{2}(x^{2}+y^{2})\big]\psi _{1}   \notag\\
&&+\big[(\alpha _{n}-\alpha _{s})\left\vert \psi _{-1}\right\vert ^{2}+(\alpha
_{n}+\alpha _{s})(\left\vert \psi _{0}\right\vert ^{2}+\left\vert \psi
_{1}\right\vert ^{2})\big]\psi _{1}  \notag \\
&&+\alpha _{s}\psi _{-1}^{\ast }\psi _{0}^{2},  \notag \\
i\frac{\partial \psi _{0}}{\partial t} &=&\big[-\frac{1}{2}(\frac{\partial^2}{\partial x^2}+
\frac{\partial^2}{\partial y^2})+\frac{1}{2%
}\Omega ^{2}(x^{2}+y^{2})\big]\psi _{0}  \notag \\
&&+\big[(\alpha _{n}+\alpha _{s})(\left\vert \psi _{1}\right\vert
^{2}+\left\vert \psi _{-1}\right\vert ^{2})+\alpha _{n}\left\vert \psi
_{0}\right\vert ^{2}\big]\psi _{0} \notag\\
&&+2\alpha _{s}\psi _{0}^{\ast }\psi _{-1}\psi _{1},  \notag \\
\!i\frac{\partial \psi _{-1}}{\partial t} &=&\!\big[-\frac{1}{2}(\frac{\partial^2}{\partial x^2}+
\frac{\partial^2}{\partial y^2})+\!\frac{1}{%
2}\!\Omega ^{2}\!(x^{2}+y^{2})\!\big]\psi _{-1}  \notag \\
&&+\big[(\alpha _{n}+\alpha _{s})(\left\vert \psi _{0}\right\vert
^{2}+\!\left\vert \!\psi _{-1}\!\right\vert ^{2})+\!(\alpha _{n}-\alpha
_{s})\!\left\vert \!\psi _{1}\!\right\vert ^{2}\!\big]\!\psi _{-1}  \notag \\
&&+\!\alpha _{s}\!\psi _{1}^{\ast }\!\psi _{0}^{2},  \label{2}\label{GP2}
\end{eqnarray}
\end{small}

\noindent
where $\alpha _{n}=N\sqrt{M^3\omega_z/(2\pi\hbar^5)}c_0$, $\alpha _{s}=N\sqrt{M^3\omega_z/(2\pi\hbar^5)}c_2$ and $\Omega=\omega_x/\omega_z$.
The particle number conservation condition becomes $\int\sum_m |\psi_m|^2 d\vec{r}=1$ $(m=0,\pm1)$.
The time and length are measured in units of $\omega_z^{-1}$ and $\sqrt{\hbar/(M\omega_z)}$, respectively.
Based on the 2D GPEs (\ref{GP2}), we are going to shade light on the dynamical properties of ring dark
solitons in spin-$1$ BEC.

Experimentally, the matter wave dark solitons can be created by means of various methods, such as
phase engineering, density engineering and quantum-state engineering \cite{Carr,Burger,Dobrek,
Denschlag,Dutton,Ginsberg,Wright}. Here we study the dynamics of ring dark solitons by simulating the phase engineering technique \cite{Burger,Dobrek,Denschlag,Carr}.

We suppose that $\omega_{\perp}=\omega_x = \omega_y = (2\pi)\times12$ Hz and $\omega_z = (2\pi)\times372$
Hz, respectively. As in \cite{Ho}, we choose $a_0=46a_B$, $a_2=52a_B$ ($a_B$ is the Bohr radius)
for $^{23}$Na condensate and $a_0=110a_B$, $a_2=107a_B$ for $^{87}$Rb condensate. And we numerically trace the
dynamics of the ring dark solitons in spin-$1$ $^{23}$Na and $^{87}$Rb condensates.

Generally, the phase engineering technique consists of passing a far-off resonant laser pulse through an absorption plate which is used to modulate the spacial distribution of the intensity of laser beam, and creating the
corresponding conservative Stark-shift potential that leads to a space-dependent phase shift in the condensate
order parameter. Fig. \ref{fig1} shows the schematic for the generation of ring dark solitons by phase engineering technique. The laser beam that is applied onto the condensate supplies an effective potential, which would produce phase distributions in the condensate order parameter. The laser beam is pulsed on for a time $\delta t$, which is much smaller than the correlation time of the condensate. The phase factor imprinted on the condensate can be adjusted by varying $\delta t$. We use the following model to simulate the phase distribution after illustration of the laser beam,

\begin{equation}
\phi(x,y) = \frac{\triangle\phi }{2}\times\big[1+{\rm tanh}(\frac{R_0-\sqrt{x^2+y^2}}{\Delta x})\big], \label{eq3}
\end{equation}

\noindent
where $\Delta\phi$, $R_0$ are the total phase difference and the radius of ring dark soliton, respectively.
$\Delta x$ reveals the width of potential edge (induced by the laser beam).

\begin{figure}[tbp]
\includegraphics[width=8cm,height=2cm]{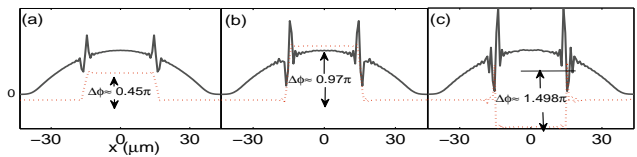}
\caption{(Color online) The 1D density $|\psi_1(x,y=0)|^2$ (solid line) and phase distribution
$\phi(x,y=0)$ (dotted line) at
$t=0.21$ $ms$ in the $^{87}$Rb condensate with $N_{Rb}=2.5\times10^4$, $\Delta x=0.54$ $\mu m$,
$R_0=15.77$ $\mu m$ and
(a) $\Delta\phi=\pi/2$,
(b) $\Delta\phi=\pi$, (c) $\Delta\phi=3\pi/2$, respectively.}
\label{fig3}
\end{figure}

\section{Structures of ring dark solitons}\label{sec3}
The ground state is antiferromagnetic for the $^{23}$Na condensate and ferromagnetic for the $^{87}$Rb condensate. Because all spinors related to each other by gauge transformation $e^{i\theta}$ and spin rotations $U(\alpha, \beta, \gamma)$ ($\alpha, \beta, \gamma$ are Euler angles) are degenerate \cite{Ho}, the ground
vectorial order parameter $(\Phi_1, \Phi_{0}, \Phi_{-1})$ \cite{Bao} in our numerical results is only one of them.
The ground state of the $^{23}$Na condensate has not the component with magnetic quantum number $m=0$. In certain cases we use ground state of the $^{23}$Na condensate that is produced by rotating the numerically obtained ground state in the spin-$1$ space. We choose the rotating operation as $U(0,\pi/2,\pi/4)$ to produce number balanced states in $\Phi_1$ and $\Phi_{-1}$ components.

\begin{figure}[tbp]
\includegraphics[width=7cm]{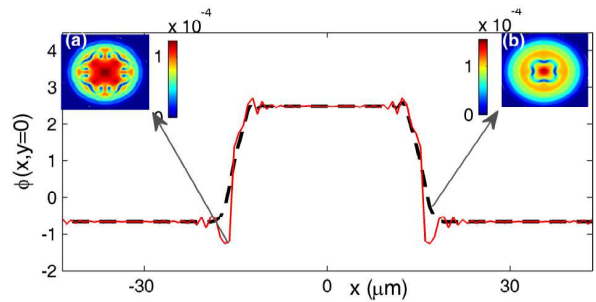}
\caption{(Color online) The 1D phase distribution $\phi(x,y=0)$ at $t=0.43$ $ms$ in
the $\psi_1$ component of the $^{87}$Rb
condensate with parameters $R_0=15.77$ $\mu m$, $\Delta\phi=\pi$, $\Delta x=1.09$ $\mu m$
(dashed line) and
$\Delta x=0.25$ $\mu m$ (solid line). (a) and (b) show the corresponding different density
decay profiles of the ring
dark solitons at $t=12.84$ $ms$ and $t=57.76$ $ms$, respectively.}
 \label{fig4}
\end{figure}

After the illustration of the far-off resonant laser pulse, the matter wave function can be written as
\begin{equation}
\Psi=(\psi_{1}, \psi_{0}, \psi_{-1})^T=(\Phi_1, \Phi_{0}, \Phi_{-1})^Te^{i\phi(x,y)},
\label{eq4}
\end{equation}

\noindent
where $\phi(x,y)$ is the phase distribution after the illumination of the laser beam as modeled
in equation (\ref{eq3}).

As a far-off resonant laser pulse passes through the condensate, all the three components of the
condensate feel the same conservative Stark-shift potential, and ring dark solitons in different components
come up with the same configuration. So we just show one component to exhibit the early stages of
the dynamics of the condensate. Using initial conditions $\Psi$ in equation (\ref{eq4}) with the numerical
ground state before the rotation in the spin space, we show the generation process of the ring dark solitons in the $\psi_1$ component of $^{23}$Na condensate in Fig. \ref{fig2} with a total number of atoms $N_{Na}=10^5$, $\Delta\phi=\pi$, $\Delta x=0.54 $ $\mu m$ and $R_0=15.77$ $\mu m$. As the total phase difference is imprinted onto the condensate, the density in each component begins to redistribute, and one obvious ring dark soliton appears, as shown in Fig. \ref{fig2} (a). At about $t=0.02$ $ms$, the ring dark soliton begins to appear, and a bright one is forming up
simultaneously inside the dark ring. Actually, the relative position of the dark and bright ring is related to the phase gradient properties, as found in \cite{Burger}.

\begin{figure}[tbp]
\includegraphics[width=7.8cm]{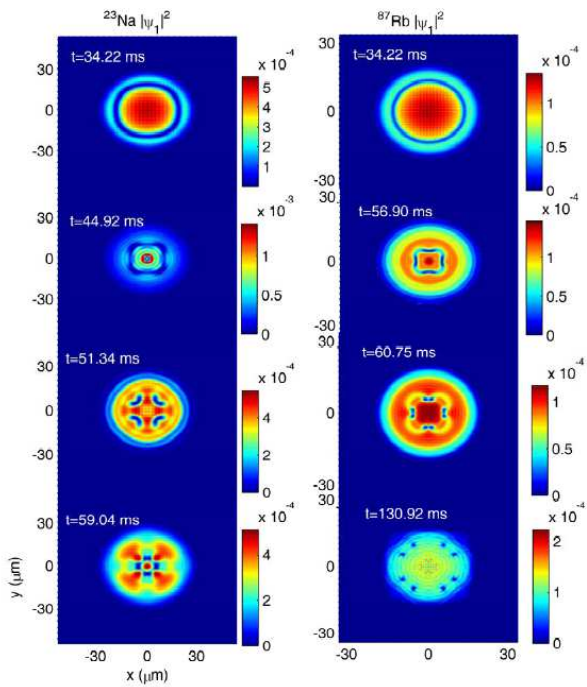}
\caption{(Color online) The density distribution $|\psi_1|^2$ of the decaying ring dark solitons
in the $^{23}$Na and $^{87}$Rb condensates with parameters $N_{Na}=N_{Rb}=2.5\times 10^4$,
$R_0=8.16$ $\mu m$, $\Delta\phi=1.05\pi$ and $\Delta x=0.56$ $\mu m$. Other components
of the condensate exhibit the same dynamic configuration. The initial condition is
modeled by equation (\ref{eq4}) with the numerical ground state before rotation in the spin space.}
\label{fig5}
\end{figure}

There are two key parameters in the phase engineering technique,  $\Delta x$ which reveals the width of the potential edge and the phase difference $\Delta \phi$. As the numerical results in Fig. \ref{fig3} show, the phase difference at $t=0.21$ $ms$ is smaller than the corresponding initial value. In fact, the phase gradient increases the total energy of the system, and the depleted phase differences are compensated in the form of the fringes of the density distribution as we can see from the Fig. \ref{fig3} (b) and (c). When $\Delta\phi=3\pi/2$, a second much shallower ring dark soliton appears, as shown in Fig. \ref{fig3} (c). For (a) and (b) in Fig. \ref{fig3}, the particle density in the trough chooses a finite value (meaning not a black one), whereas the generated ring dark soliton takes a nearly black form with respect to the density characteristics in Fig. \ref{fig3} (c). When we further increase the total phase difference, such as
$\Delta\phi=2\pi$, $3\pi$, more ring dark solitons will come up and show collective oscillations.

Next, we aim to find out the effects of the Stark-shift potential edge width on the dynamics of
the ring dark solions through varying $\Delta x$. As $\Delta x$ varies from $ 2.17 $ $\mu m$ to $0.25 $ $\mu m$,
the generated ring soliton gets darker and darker. But when $\Delta x<0.54$ $\mu m$, the lifetime of ring
dark soliton decreases fiercely and the ring dark soliton shows different decay configurations. Using initial condition
modeled in equation (\ref{eq4}) for the $^{87}$Rb condensate, Fig. \ref{fig4} shows the phase of the $\psi_1$ component in $x$ direction with the dashed (solid) line corresponding to $\Delta x=1.09$ $\mu m$ ($0.25$ $\mu m$). We can see that the phase jump is nearly the same, except for small amplitude oscillations for the $\Delta x=0.25$ $\mu m$ case. These two cases correspond to different decay properties of ring dark solions. For $\Delta x=0.25$ $\mu m$, the
lifetime of the ring dark solions is shorter, and eight vortex and anti-vortex pairs are produced. But for
$\Delta x=1.09$ $\mu m$, the ring dark soliton can last as long as $57.76$ $ms$, and finally decays into four vortex and anti-vortex pairs. This can be understood by resorting to the healing length $\xi=1/{\sqrt{8 \pi an}}$ \cite{Pethick}.
The size of topological excitations such as dark solitons and vortices is fixed by the healing length \cite{Pethick}. And for the $^{87}$Rb condensate in our system, $\xi\approx 0.55$ $\mu m$, so it is reasonable to assume the unstability of the soliton structures with $\Delta x <0.55$ $ms$. In Fig. \ref{fig4}, the notable oscillation of the phase distributions around the slope enhances the decay of the ring structures.

\begin{figure}[hbp]
\includegraphics[width=8cm,height=4cm]{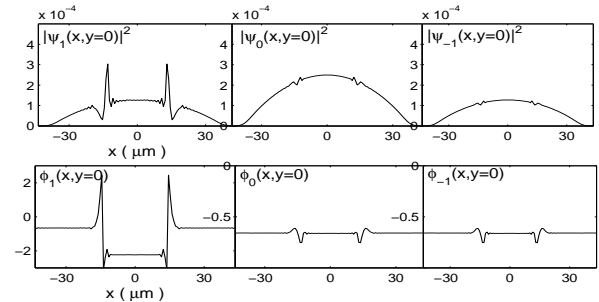}
\caption{The 1D density (top panels) and phase (bottom panels) distribution at
$t=0.43$ $ms$ for the three components
of the $^{87}$Rb condensate, with $N_{Rb}=2.5\times10^4$, $R_0=15.77$ $\mu m$,
$\Delta\phi=3\pi/2$ and $\Delta x=1.09$ $\mu m$.}
\label{fig6}
\end{figure}

\begin{table}[htb]\renewcommand{\arraystretch}{1.5}
\caption{lifetime of the ring dark solitons in different BEC systems. For the scalar BEC with Feshbach resonance management (FRM), the total phase difference is $\Delta\phi=0.707\pi$, which should be a key factor for long-lived ring dark soliton. For the spinor $^{23}$Na or $^{87}$Rb BEC case, ring dark solitons with total phase difference $\Delta\phi\approx\pi$ are generated in each component of the condensate.
}
\label{table1}
\centering
\begin{tabular}{lc}
\par
\hline\hline
Systems of BEC & Lifetime (units of $1/\omega_{\perp}$)\\
\hline
Scalar BEC without FRM & $1.14$ \\
Scalar BEC with FRM &  $5.09$ \\
Spinor $^{23}$Na BEC without FRM & $3.48$ \\
Spinor $^{87}$Rb BEC without FRM & $4.30$ \\
\hline\hline
\end{tabular}
\end{table}

To explore the differences of the dynamics of the ring dark solitons between $^{23}$Na and
$^{87}$Rb condensates, we plot the density distributions of the decaying ring dark solitons. As shown in Fig. \ref{fig5}, the density distribution of the $\psi_1$ component of $^{23}$Na and $^{87}$Rb condensates exhibits
the different decaying dynamics. At $t=34.22$ $ms$, nearly black ring dark solitons have formed
in both condensates. However, ring dark solitons in the $^{23}$Na, which break into x-like
vortex pairs, live a shorter lifetime comparing to that in the $^{87}$Rb condensate which break
into +-like vortex pairs. The dynamics of the vortex pairs is the same as found by \cite{Theocharis}
in scalar BEC. The configuration of the vortex and anti-vortex pairs oscillates between the x- and
+-like configurations, while the ring periodically shrinks and expands.

If all the components of the condensate bear the same parameterized ring dark solitons, the total density distribution of the condensate exhibits the same profile as that in each component. When detected in the experiment with time of flight technique, the profiles of ring dark solitons do not show differences from that in scalar BEC.
In Table \ref{table1}, we summarize the lifetimes of ring dark solitons in different systems of BEC according to references \cite{Theocharis, Hu} and our numerical calculations. We note that with certain parameters (as shown in Fig. \ref{fig5}) in $^{87}$Rb condensate, the lifetime of the nearly black ring soliton in spinor BEC without Feshbach resonance management (FRM) is remarkably prolonged referring to the lifetime of ring dark soliton in scalar BEC. And this lifetime scale is comparable with the lifetime of a shallower ring dark soliton (the total phase difference is $\Delta\phi=0.707\pi$) in \cite{Hu}. It would be useful to explore ways to prolong the lifetime of the ring dark solitons by varying the parameters of the external potentials or the spin-dependent and independent interactions.

\begin{figure}[tbp]
\includegraphics[width=8cm]{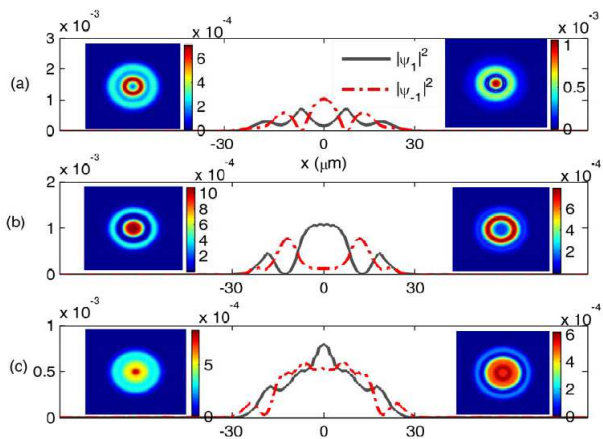}
\caption{(Color online) The 1D density distribution at (a) $t=67.60$ $ms$ (b) $t=121.50$ $ms$ and (c)
$t=159.15$ $ms$ in $\psi_1$ (solid), and $\psi_{-1}$ (dash-dotted) components of the
$^{23}$Na condensate showing the ``darkness-transferring" effect with
$N_{Na}=2.5\times 10^4$, $R_0=8.16$ $\mu m$, $\Delta\phi=1.05\pi$ and $\Delta x=0.56$ $\mu m$.
The 2D density distributions
according to each time point are attached for $\psi_1$ (left) and $\psi_{-1}$ (right). The density
of $\psi_{0}$ component (which is of the order of $10^{-8}$) is omitted.}
\label{fig7}
\end{figure}

\section{induced and off-centered ring dark solitons }\label{sec4}
\subsection{Common characteristics of the coexistence state of ring dark solitons in $^{23}$Na and $^{87}$Rb condensates}
In real experiments, the far-off resonant laser pulse is not ideal, so we consider symmetry breaking effects in
our simulation by allowing the trap center to randomly jump within a region $[-\delta, \delta] \times [-\delta, \delta]$ ($\delta = 0.002h$, where $h$ is the grid size).

We suppose that the condensate takes phase patterns as
\begin{equation}
\Psi=(\psi_{1}, \psi_{0}, \psi_{-1})^T=(\Phi_1e^{i\phi(x,y)}, \Phi_{0}, \Phi_{-1})^T.
\label{eq5}
\end{equation}
On this occasion, the ring dark soliton is formed in the first component, and at the same time,
ring dark solitons in other components are induced. And the mutual filling of different components
supplies effective potentials to support the ring dark soliton structure in each component, resulting in a
coexistence structure which can live hundreds of milliseconds.

\begin{figure}[tbp]
\includegraphics[width=9.5cm,height=3.7cm]{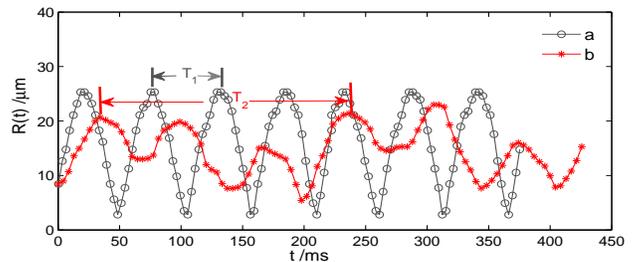}
\caption{(Color online) The time evolution of the radius of the ring dark solitons in $^{23}$Na condensate with $N_{Na}=2.5\times 10^4$, $R_0=8.16$ $\mu m$ and $\Delta x=0.56$ $\mu m$. Line a: radius of ring dark solitons produced by generating same parameterized ring dark solitons in each component of the condensate with an initial total phase difference $\Delta\phi=0.5\pi$; line b: radius of ring dark solitons produced by generating one ring dark soliton in one component of the condensate with an initial total phase difference $\Delta\phi=1.05\pi$.}
\label{fig8}
\end{figure}

In Fig. \ref{fig6}, we plot the $1$D density and phase distributions of the three components at $t=0.43$ $ms$ for
the $^{87}$Rb condensate. We find that the ring dark soliton in the $\psi_1$ component has been formed while
another two much shallower dark solitons in the $\psi_0$ and $\psi_{-1}$ components germinate as slight phase
jumps are in their way (see the bottom panel). In fact, the atoms in the $\psi_0$ and $\psi_{-1}$ components fill the density gap in the $\psi_1$ component, resulting in a complex interdependent structure. After the full generation of ring dark solitons in all the three components, the coexistence structure oscillates complicatedly.

In order to exhibit some aspects of the complex oscillating properties, we depict the darkness of the solitons in
different components (the velocity of the dark soliton is connected to the darkness) \cite{Jackson}. In our numerical
calculation, both $^{23}$Na and $^{87}$Rb condensates exhibit a darkness-transferring property between different components. Fig. \ref{fig7} shows the dynamical process of the transferring for the $^{23}$Na condensate using an initial condition modeled by equation (\ref{eq5}) with ground state that directly obtained by imaginary time evolution. At $t=67.60$ $ms$ the nearly black ring soliton appears in the $\psi_{-1}$ component. And by $t=121.50$ $ms$, the darkness of the ring soliton in the $\psi_{-1}$ component has totally been transferred onto the $\psi_{1}$ component (as shown in Fig. \ref{fig7} (b)). After $37.65$ $ms$ the ring dark soliton restores its darkness in the $\psi_1$ component as shown in Fig. \ref{fig7} (c). For the phase distribution model in equation (\ref{eq5}) and parameters in Fig. \ref{fig7}, the coexistence structure in both $^{23}$Na and $^{87}$Rb condensates can live more than $800$ $ms$.

Even though complicated, the oscillations exhibit periodicity roughly for an initial condition with
the absence of the $m=0$ component. And the oscillation frequency is approximately $0.58\Omega/\sqrt{2}$, which is different from that in the scalar BEC \cite{Theocharis, Konotop}. The periodical signature can be seen from Fig. \ref{fig8}, showing the evolutionary radius of the ring dark soliton. In Fig. 8, line $a$ corresponds to the state of generating same ring dark solitons in all the three components of the condensate, while line $b$ the state of generating only one obvious ring dark soliton in one component of the condensate. For line $a$, the oscillation frequency is  $\omega_o\approx 2.14\Omega/\sqrt{2}$ (roughly consistent with that in \cite{Theocharis} for the scalar BEC) while it is $\omega_o\approx0.58\Omega/\sqrt{2}$ for line $b$, indicating a prolonged oscillation period. In contrast with the $1$D condensate, ring dark soliton oscillates between the maximum and minimum radius with nearly a two times bigger frequency in the $2$D condensate. For the coexistence state composed of interdependent ring dark solitons coming from different components of the condensate, the oscillation exhibit periodic character in a rough way because of the interaction and radiation of the ring dark solitons. During the rough oscillation process, a prolonged oscillation period is found. In fact, the oscillation of the ring dark solitons is hindered by the embedded particles of another component, resulting in a low oscillation frequency of the ring dark solitons. We should mention that if we use initial condition with the appearance of all the three components of the condensate, the oscillation is more complicated and so is the periodicity.

\begin{figure}[tbp]
\includegraphics[width=7.5cm]{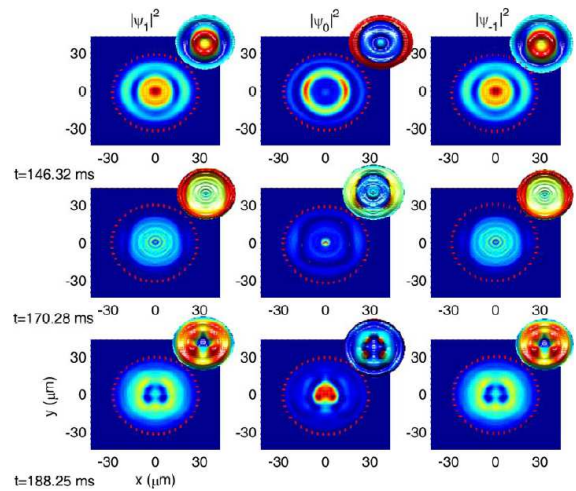}
\caption{(Color online) The decaying process shown by the density and phase
distributions of the $^{23}$Na condensate with an initial
off-centered ring dark soliton in the $\psi_0$ component. The deviation in $y$
direction $\delta p$ is $0.6 $ $\mu m$ and $R_0=8.16$ $\mu m$,
$\Delta\phi=3\pi/2$, $\Delta x=0.56$ $\mu m$. At $t=146.32$ $ms$, imbalance in
the ring dark solitons appears, and $24.96$ $ms$ later, the ring dark soliton distorts in the $\psi_0$ component. The ring dark solitons are
totally broken with vortex and anti-vortex pairs in the $\psi_{0}$ component at $t=188.25$ $ms$.}
\label{fig9}
\end{figure}

If the coexistence state is induced by an obviously off-centered ring dark soliton,
it suffers an earlier collapse for both $^{23}$Na and $^{87}$Rb condensates with a time scale
of about $150$ $ms$. In the following, taking into account the fact that the center of the ring may not be coincide
with that of the trap, we suppose an off-centered ring dark soliton with an deviation $\delta p$.
The oscillation of the ring dark solitons are sensitive to such a deviation. When the deviation increases to the value $\delta p$ =$0.20$ $ms$, the lifetime of the ring dark solitons decreases evidently. In Fig. \ref{fig9}, we show the decaying process of the ring dark solitons with $\delta p$ = $0.60 $ $\mu m$ for $^{23}$Na. The parameters are the same as that in Fig. \ref{fig11} (a)-(c). Without deviation $\delta p$, the coexistence state of ring dark solitons can live as long as $320.88$ $ms$ and breaks into four pieces in the $\psi_0$ component at $t=378.64$ $ms$ without the formulation of vortex and anti-vortex pairs as shown in Fig. \ref{fig11} (a)-(c). With a finite $\delta p$ the coexistence state suffers breakdown at a much earlier time.  At $t=146.32$ $ms$, the darkness of the rings
loose their balance, and the darker part of the ring is more inclined to decay into vortices.
At $t=170.28$ $ms$, the ring structure in the $\psi_{0}$ component distorts slightly into a form of trapezoid and four vortex and anti-vortex pairs come up , as shown by the corresponding top-right phase profile panel. And at $t=285.50$ $ms$, the ring dark solitons break down totally. The numerical results show that the lifetime and dynamics of the ring dark solitons are sensitive to such an deviation, indicating that it is very essential in the real experiment to control the relative position of the optical potential and the far-off resonant laser beam which is used to generate the ring dark solitons.

\begin{figure}[tbp]
\includegraphics[width=7cm]{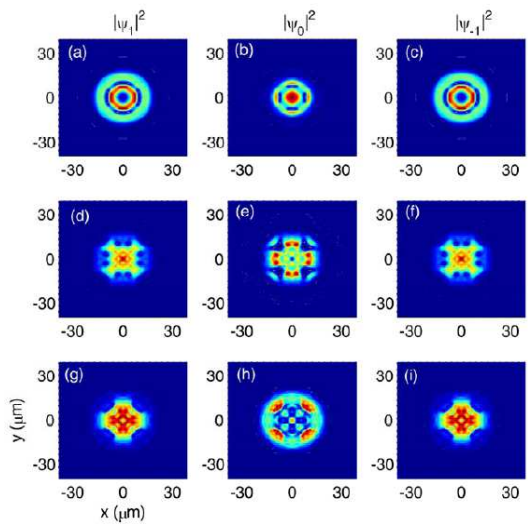}
\caption{(Color online) The density distributions $|\psi_1|^2$, $|\psi_0|^2$ and $|\psi_{-1}|^2$
of $^{87}$Rb condensate at chosen time points with $N_{Rb}=2.5\times 10^4$, $R_0=8.16$ $\mu m$,
$\Delta\phi=3\pi/2$ and $\Delta x=0.56$ $\mu m$, showing the decay of the ring dark solitons
and motion of the vortex and anti-vortex pairs. (a)-(c), (d)-(f) and (g)-(i) are density
configurations at $t=123.22$ $ms$, $t=164.28$ $ms$, and $t=189.10$ $ms$, respectively.}
\label{fig10}
\end{figure}

\subsection{Different aspects of the coexistence state of ring dark solitons in $^{23}$Na and $^{87}$Rb condensates}
To explore the differences between $^{23}$Na and $^{87}$Rb condensates in a short time scale, we suppose the initial
vector parameter as

\begin{equation}
\Psi=(\psi_{1}, \psi_{0}, \psi_{-1})^T=(\Phi_1, \Phi_{0}e^{i\phi(x,y)}, \Phi_{-1})^T,
\label{eq6}
\end{equation}
with the parameters $N_{Na}=N_{Rb}=2.5\times 10^4$, $R_0=8.16$ $\mu m$, $\Delta\phi=3\pi/2$ and $\Delta x=0.56$ $\mu m$. Here, for the $^{23}$Na condensate, the ground state is obtained by the acting of $U(0, \pi/2, \pi/4)$ on the ground state that we obtain from direct numerical calculation.

\begin{figure}[tbp]
\includegraphics[width=7.8cm,height=5cm]{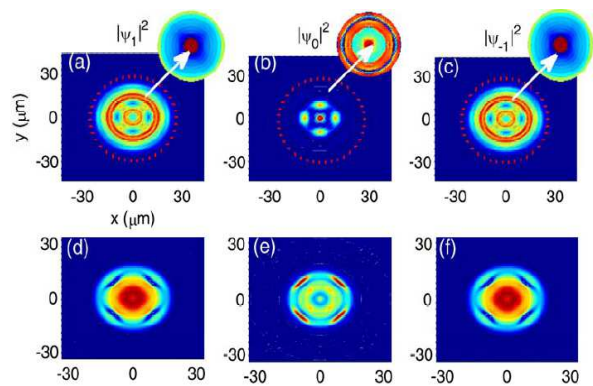}
\caption{(Color online) (a)-(c) are the density distributions $|\psi_1|^2$, $|\psi_0|^2$ and $|\psi_{-1}|^2$
of $^{23}$Na condensate at $t=378.64$ with $N_{Na}=2.5\times 10^4$, $R_0=8.16$ $\mu m$, $\Delta\phi=3\pi/2$
and $\Delta x=0.56$ $\mu m$. The insets on the top right shows the phase distribution of the condensate.
(d)-(f) are density distributions at $t=160.00$ $ms$ with the same parameters as in Fig. \ref{fig10} except for
$\alpha_s=0.038\alpha_n$. The ground sate used have been rotated by $U(0,\pi/2,\pi/4)$ in the spin space.}
\label{fig11}
\end{figure}

In this case, the coexistence state structure decays into vortex and anti-vortex pairs at an earlier time
for the $^{87}$Rb condensate. As it is shown in Fig. \ref{fig10}, at about $t=123.22$ $ms$, four vortex
and anti-vortex pairs arranging themselves in a ``+" configuration have been formed in the $\psi_0$ component of the
condensate while vortex pairs are emerging in the other two components. The vortex and
anti-vortex pairs afterwards arrange themselves in +-like and x-like configurations alternatively.
Following the vortex configuration in the $\psi_0$ component, the vortex and anti-vortex pairs take the configuration of ``+" in $\psi_1$ and $\psi_{-1}$ components at $t=164.28$ $ms$, while they take the configuration of ``x" in the $\psi_0$ component. At $t=189.10$ $ms$, the vortex and anti-vortex pairs in $\psi_1$ and $\psi_{-1}$ components catch up with that in the $\psi_0$ component, and evolve into x-like configurations. However, this kind of vortex pairs oscillations do not occur in the $^{23}$Na condensate with a time span of $378.64$ $ms$, at which the ring dark solitons breaks into four pieces in the $\psi_0$ component as shown in Fig. \ref{fig11} (a)-(c). The insets to the top-right of the density distributions show the phase distributions, showing the absence of the vortex and anti-vortex pairs.
The coexistence state composed of ring dark solitons coming from different components in $^{23}$Na condensate enjoy a longer lifetime than that in the $^{87}$Rb condensate. We ascribe this phenomena to the effect of the spin-dependent interaction parameter $c_2$, which is positive for $^{23}$Na and negative for $^{87}$Rb.

\begin{figure}[tbp]
\includegraphics[width=9.5cm]{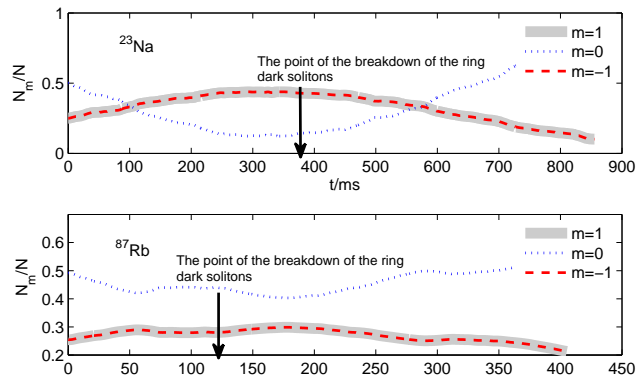}
\caption{(Color online) Spin populations versus evolution time of $^{23}$Na (upper panel) and $^{87}$Rb (bottom panel) condensates with $N_{Na}=N_{Rb}=2.5\times 10^4$, $R_0=8.16$ $\mu m$, $\Delta\phi=3/2\pi$ and $\Delta x=0.56$ $\mu m$. For the $^{23}$Na condensate, a rotated initial ground state is used.}
\label{fig12}
\end{figure}

Taking the same parameters in the $^{87}$Rb condensate but with a positive $\alpha_s$ ($\alpha_s = 0.038\alpha_n$ is chosen in our numerical calculation), we find that the lifetime of the coexistence state composed of ring dark solitons is prolonged remarkably, as show in Fig. \ref{fig11} (d)-(f). In the real $^{87}$Rb condensate with parameters in Fig. \ref{fig10}, the coexistence sate breaks into vortex and anti-vortex pairs totally at $123.22$ $ms$ (Fig. \ref{fig10} (a)-(c)). However, the state can last until $t=160.86$ $ms$ as shown in Fig. \ref{fig11} (d)-(f).  Different spin mixing dynamics which is determined by the spin-dependent interaction term are found between $^{23}$Na and $^{87}$Rb condensates as shown in Fig. \ref{fig12}, and this should be responsible for the different lifetimes of the ring dark solitons between $^{23}$Na and $^{87}$Rb condensates.

So ring dark solitons of the ferromagnetic condensate incline to decay into vortex and anti-vortex pairs much earlier than that of the antiferromagnetic condensate. As it is known, modulational instability is possible in ferromagnetic spinor condensates \cite{Nistazakis2008, Nicholas}. And the instability would cause the breakdown of exotic structures such as ring dark solitons. Calculations in the present paper show that such a worry is not necessary. The modulational instability has a close dependence on the phase distribution of the state, as found in \cite{Nicholas}. By simulating the experimental generation of ring dark solitons, the state in our present setup, which is composed of interdependent ring dark solitons, is not a stationary state. So the condition for the modulational instability in \cite{Nistazakis2008, Nicholas} is not well satisfied. We can understand this point by referring to the spin-mixing effect. States that would suffer modulational instability found in \cite{Nistazakis2008, Nicholas} do not involve the spin-mixing \cite{Chang} between components of the condensate. In our system, spin-mixing occurs at the very beginning as shown in Fig. \ref{fig12}, which exhibits the spin populations versus evolution time (spin-mixing in other parameter region has also been found), indicating the deviation of the state of ring dark solitons from the modulationally instable states obtained in \cite{Nistazakis2008, Nicholas}.

\section{CONCLUSIONS} \label{sec5}
In conclusion, the generation and dynamics of the ring dark solitons in both spin-$1$ $^{23}$Na and $^{87}$Rb condensates are explored by simulating the phase engineering technique numerically. We find that the characteristics of
the far-off resonant laser pulse, such as the Stark-shift potential width (revealed by $\Delta x$)
and the duration of the beam $\delta t$ (revealed by $\Delta \phi$), affect the dynamical properties
of the condensate dramatically, including the lifetime and the decay profiles of the ring dark solitons.
We choose optimal parameters to generate nearly black ring solitons. If only one ring dark soliton in one
component of the condensate is generated, ring dark solitons in other components are induced, and they compose a coexistence state, which exhibits dynamical oscillations for hundreds of milliseconds. The oscillation of the coexistence state can live hundreds of milliseconds because of the mutual filling and spin mixing of different
components. The coexistence state is more likely to be stabilized in presence of all the
three components of the condensate. The coexistence state can enjoy a longer lifetime in the $^{23}$Na condensate than
that in the $^{87}$Rb condensate. However, if the coexistence structure is induced by an obviously off-centered
ring dark soliton, it suffers an earlier collapse with the appearance of vortex and anti-vortex pairs. This work may help to realize manipulation of ring dark solitons in high-dimensional multicomponent BECs. Because of the long lifetime of the ring structures (more than $800$ $ms$ for $^{87}$Rb), we hope that our results can stimulate investigation of
high-dimensional ring dark solitons in future spinor BEC experiments. In fact, the density engineering and phase engineering can be utilized together to generate more perfect ring dark structures, whose lifetime should be much longer. Besides, the FR technique is waiting to be used to manipulate ring dark solitons in spinor BEC in the future.

\textbf{Acknowledgments}
This work was supported by the NKBRSFC under grants Nos. 2011CB921502, 2012CB821305, 2009CB930701, 2010CB922904, NSFC under grants Nos. 10934010, 60978019, and NSFC-RGC under grants Nos. 11061160490 and 1386-N-HKU748/10. D. S. Wang was supported by NSFC under grant No. 11001263 and China Postdoctoral Science Foundation. H. Wang was support by NSFC under grant No. 10901134.

\end{document}